\documentclass[osajnl,twocolumn,showpacs,superscriptaddress,10pt]{revtex4-1} 
\usepackage{amsmath,amssymb,graphicx}
\usepackage{float}
\usepackage{hyperref}  
\newcommand{\ket}[1]{{|#1\rangle}}
\newcommand{\bra}[1]{{\langle#1|}}

\newcommand{\eq}[1]{{eq.(\ref{#1})}}
\DeclareMathOperator{\Tr}{Tr}

\begin{document}

\title{Majorization and Measures of Classical Polarization in Three Dimensions}

\author{Omar Gamel}\email{Corresponding author: ogamel@physics.utoronto.ca}
\author{Daniel F. V. James}
\affiliation{Department of Physics, University of Toronto, 60 St. George St., Toronto, Ontario, Canada M5S 1A7.}

\begin{abstract}
There has been much discussion in the literature about rival measures of classical polarization in three dimensions. We gather and compare the various proposed measures of polarization, creating a geometric representation of the polarization state space in the process. We use majorization, previously used in quantum information, as a criterion to establish a partial ordering on the polarization state space. Using this criterion and other considerations, the most useful polarization measure in three dimensions is found to be one dependent on the Bloch vector decomposition of the polarization matrix.
\end{abstract}


\maketitle 

\section{Introduction}

The polarization state of a classical beam of light has been quantified and studied since the nineteenth century, pioneered by the well-known model of the Stokes vector and its geometric representation in the Poincar\'{e} sphere \cite{stokes, poincare}. Modern mathematical models of polarization in statistical optics rely more heavily on the polarization matrix, but maintain the same underlying physical picture of an electric field in two dimensions perpendicular to the direction of propagation \cite{wolf07}. 

Based on either the Stokes vector or the Polarization matrix, there emerges a natural unambiguous definition of the degree of polarization of such a classical beam with a two dimensional electric field, based on the fraction of total power that is is in the polarized component \cite{qasimiwolf}. 

However, if we extend our inquiry to a three dimensional electric field distribution, we find there is no single unambiguous generalization of the degree of polarization. This has led to many authors suggesting different measures of classical polarization in three dimensions. Some measures are based on  quantum purity \cite{nielsenchuang}, a decomposition of the polarization matrix in $SU(3)$ \cite{bergman, friberg02}, non-quantum entanglement \cite{qianeberly}, von Neumann entropy \cite{gil07}, the fully polarized field component \cite{wolf04}, and the invariants of the rotational group \cite{barakat83}.

In this paper, we add to the analysis in our previous publication \cite{gamelpurity} by comparing the measures above. Here we use {\it majorization} as a key comparison criterion. Majorization is a technique to partially order real vectors, and has been used extensively in quantum information to place a partial ordering on the degree of entanglement of bipartite states \cite{nielsenmajorization}. Here we show that it can be extended to create a partial ordering on the state of polarization in three dimensions (or higher). 


In section \ref{s2:definemeasures} we define the different measures of three dimensional polarization, and illustrate the relationship between them. Section \ref{s3:majorization} discusses majorization and the requirement that the degree of polarization be a Schur-convex function. Section \ref{s4:statespace} explores the state space of polarization states in terms of relevant eigenvalues, and provides a graphical representation. Section \ref{s5:majtest} applies the criteria in the previous two sections to the measures we have defined, to find which of them respect the criterion and which can be discarded. Section \ref{s6:graphcomp} outlines all the measures graphically by providing contours of equal polarization. Some considerations based on these contour plots are used to select the most versatile measure of polarization. Finally, section \ref{s7:Ndimensions} generalizes the results to $N$ dimensions.

Our analysis adds to and clarifies much of the discussion on measures of higher dimensional polarization in the literature \cite{luis05, friberg09, friberg10, sheppard, gil07, gil10}, classifying the measures and analyzing their relationship. 


\section{Measures of Polarization in Three Dimensions}\label{s2:definemeasures}

Consider an electric field distribution in three dimensions. The complex electric field values in the $x$, $y$ and $z$ directions are taken to be probabilistic ensembles given by $E_1$, $E_2$ and $E_3$ respectively. 

The polarization state of the beam of light is given by the 3$\times$3 polarization matrix $\Phi$, defined as
\begin{equation}
\Phi_{ij}^{(3)}=\langle E_iE_j^* \rangle, \hspace{20pt} i,j=1,2.3.
\label{phidef}
\end{equation} 
If one thinks of the electric fields $E_i$ as random variables, then $\Phi^{(3)}$ is their variance-covariance matrix. Note that in two dimensions, the polarization matrix $\Phi^{(2)}$ is defined the same as above, but the indices only take values 1 and 2.

The degree of polarization, $P^{(2)}$ of a two dimensional polarization matrix $\Phi^{(2)}$ is derived by writing $\Phi^{(2)}$ as the unique sum of two polarization matrices, one completely unpolarized (a multiple of the identity matrix), and one completely polarized (a rank 1 matrix) \cite{qasimiwolf,wolf07}. The standard degree of polarization in two dimensions is then the ratio of the power contained in the completely polarized matrix to the total power. It is given by
\begin{align}
P^{(2)}&=\sqrt{1-\frac{4\det(\Phi)}{\Tr[\Phi]^2}},
\label{2dpolarizationPhi}
\end{align}
where the dimensional superscript on $\Phi$ has been suppressed. 

We note that for any dimension, the degree of polarization is a basis independent property, and the matrix $\Phi$ will be positive by definition, and therefore diagonalizable. Therefore one can choose the basis which diagonalizes $\Phi$, and the degree of polarization becomes a function of the diagonal entries, i.e. the eigenvalues of $\Phi$. 

Reverting to the three dimensional case, the polarization matrix $\Phi^{(3)}$ has three non-negative eigenvalues:
\begin{align}
\lambda_1 \ge \lambda_2 \ge \lambda_3 \ge 0,  \nonumber\\
\lambda_1 + \lambda_2 + \lambda_3 = 1.
\label{lambdaconditions}
\end{align}

We haved assumed the total power is normalized, the polarization matrix has unit trace, and so the eigenvalues sum to unity. The state where $\lambda_1 = \lambda_2 = \lambda_3 = \frac{1}{3}$ is fully unpolarized, where the state where $\lambda_1 = 1$ and $\lambda_2 = \lambda_3 = 0$ is the maximally polarized state. Our measures of polarization are defined such that the fully unpolarized (mixed) state has the minimum value of polarization, and the maximally polarized (pure) state has the maximum value.

Several measures of polarization in three dimensions have been used in the literature. In the remainder of this section, we list the most prevalent measures along with a brief explanation and references that explain them in more detail. Most of the measures below have been analyzed in detail by the authors in ref. \cite{gamelpurity}. However, in this last reference, the measures of polarization were linearly rescaled to vary between 0 and 1, and different notation was used. No such rescaling is adopted in this paper. Going forward, we also suppress the dimensional superscript in the polarization matrix $\Phi$.

One can treat the polarization matrix $\Phi$ in a similar way one would treat a density matrix for a quantum system. Since classical polarization is mathematically the same as quantum purity \cite{gamelpurity}, one can define the {\it standard polarization} based on the common measure of purity of quantum systems:
\begin{equation}
P_s = \lambda_1^2 + \lambda_2^2 + \lambda_3^2.
\label{pstandard}
\end{equation}
One can also quantify the polarization by writing the polarization matrix $\Phi$ as a linear combination of some basis matrices in $SU(3)$, usually the Gell-Mann matrices \cite{gellmann}. Treating the coefficients of the basis matrices in this decomposition as components of a generalized Bloch vector \cite{bloch}, one can use the magnitude of this vector as a measure of polarization \cite{bergman,friberg02}. The vector is found to lie within an irregular region which itself lies within an eight dimensional hypersphere \cite{kimura}. The vector represents the radial distance from the sphere's origin. Therefore, we call the magnitude of the generalized Bloch vector the {\it Bloch polarization}, given by
 \begin{equation}
 P_{bl} = \sqrt{\frac{1}{2}(3P_s - 1)}.
 \label{pbloch}
 \end{equation}
One can also measure polarization making use of the Schmidt decomposition \cite{schmidt}. One first writes the electric field as an element in a product space of the one Hilbert space that represent the directional vector of the field and another representing the magnitude of this element as a function of time and space. One can then use the Schmidt decomposition to write the field as the sum of three mutually orthogonal product states, with coefficients $\kappa_1$, $\kappa_2$, and $\kappa_3$. The degree of polarization can is then provided by a weight parameter of the Schmidt decomposition, which yields the non-integer effective number of dimensions needed by the optical field \cite{qianeberly}:
\begin{equation}
K = \frac{1}{\kappa_1^4 + \kappa_2^4 + \kappa_3^4}. \nonumber
\end{equation}
By taking the outer product of the electric field (in Schmidt form) with itself, and then tracing out the function space, one is left with a $3\times 3$ matrix in the directional space, whose diagonals are $\kappa_1^2, \kappa_2^2$ and $\kappa_3^2$. Identifying this matrix with the polarization matrix $\Phi$, we see that $\lambda_i = \kappa_i^2$, for $i=1,2,3$. Therefore, based on $K$ above, we can define the {\it Schmidt polarization} as
\begin{equation}
P_{sc} = \frac{1}{P_s}. 
\label{pschmidt}
\end{equation}
This measure of polarization, uniquely among the ones considered in this paper, takes lower values for more polarized states. For example, a maximally polarized state has the minimal number of effective dimensions, 1.

Shannon entropy is also used in classical systems to quantify uncertainty about a random variable \cite{shannon}. Using this concept, one can define the {\it Von Neumann entropy} \cite{nielsenchuang}:
\begin{equation}
P_v = \lambda_1 \log_2\lambda_1 + \lambda_2 \log_2\lambda_2  + \lambda_3 \log_2\lambda_3.
\label{pvonn}
\end{equation}

If an eigenvalue $\lambda_k = 0$, we take $\lambda_k \log_2(\lambda_k) = 0$, since $\lim_{x\rightarrow 0^+} x \log(x) = 0$.

One can also write the diagonalized $3 \times 3$ polarization matrix $\Phi$ into a unique positive linear combination of the identity matrix, a rank 2 matrix with degenerate eigenvalues, and a rank 1 matrix: 
\begin{align}
(\lambda_1 {-} \lambda_2) \underbrace{\begin{bmatrix} 1 & 0 & 0 \\ 0 & 0 & 0 \\ 0 & 0 & 0 \end{bmatrix}}_{\text{rank 1}} + 
(\lambda_2 {-} \lambda_3) \underbrace{\begin{bmatrix} 1 & 0 & 0 \\ 0 & 1 & 0 \\ 0 & 0 & 0 \end{bmatrix}}_{\text{rank 2}}
&+ \lambda_3 \underbrace{\begin{bmatrix} 1 & 0 & 0 \\ 0 & 1 & 0 \\ 0 & 0 & 1 \end{bmatrix}}_{\text{rank 3}}. \nonumber
\label{rhodecompose}
\end{align}
One can take the power in the fully polarized component (i.e. the coefficient of the rank 1 matrix) as a measure of the degree of polarization \cite{wolf04}. We define the {\it order 1 polarization} or {\it full polarization} as
\begin{equation}
P_1 = \lambda_1 - \lambda_2.
\label{p1}
\end{equation}
One can also take the power in both the rank 1 and rank 2 polarized component (i.e. the sum of both coefficients) as a measure of polarization. We define the {\it order 2 polarization} as
\begin{equation}
P_{2} = \lambda_1 - \lambda_3.
\label{p2}
\end{equation}
Finally, one can make use of a hierarchy of polarization measures due to Barakat \cite{barakat83}. They are based on the coefficients of the characteristic polynomial of the polarization matrix $\Phi$. In the three dimensional case, the hierarchy has only two measures, the first of which is identical to $P_{bl}$ in \eq{pbloch} above. The second, which we call {\it Barakat's last measure}, is given by
\begin{equation}
P_{b} = \sqrt{1 - 27\lambda_1\lambda_2\lambda_3}.
\label{pbarakat}
\end{equation}
We make a few observations about the seven measures of polarization in eqs. (\ref{pstandard} - \ref{pbarakat}) above. Note that redefining them in two dimensions, $P_{bl}$, $P_1$, $P_2$ and $P_b$ will all reduce identically to $P^{(2)}$ in \eq{2dpolarizationPhi}. The measures $P_s$, $P_v$, and $P_{sc}$ will not reduce in the same way.

Also note that $P_{bl}, P_{sc}$, and $P_s$ are all functions of one another, and therefore will all induce the same ordering on the space of polarization states. We see that most purposes of comparison it suffices to use only one of them.

\section{Majorization criterion on degree of polarization}\label{s3:majorization}

The polarization state at any point should be basis indepdendent, i.e. invariant under rotation. Therefore, it is given by a basis independent property, the spectrum of eigenvalues $\vec{\lambda} = (\lambda_1, \lambda_2, \lambda_3)$ of the polarization matrix $\Phi$, satisfying the conditions in \eq{lambdaconditions}. Each of the measures of purity $P_b$, $P_v$, $P_f$ and $P_{bl}$ induces a different ordering on the space of polarization states. In this section, we introduce a physically motivated partial ordering of polarization states based on the largest and smallest eigenvalues. We then define majorization, and show that it induces the same partial ordering on the space of polarization states. We then require that our chosen measure of purity satisfy this ordering, with justification.

If one thinks of the polarization matrix $\Phi$ as a quantum density matrix, and $\ket{\psi}$ as some arbitrary pure state, the maximum possible 'overlap' with a pure state is 
\begin{equation}
\max_\psi \bra{\psi}\Phi\ket{\psi} = \lambda_{\text{max}},
\end{equation}
the maximum eigenvalue. Analogously, the minimum possible overlap is
\begin{equation}
\min_\psi \bra{\psi}\Phi\ket{\psi}=  \lambda_{\text{min}},
\end{equation}
the minimum eigenvalue.

One may argue that a larger $\lambda_{\text{max}}$ or a smaller $\lambda_{\text{min}}$ indicates a greater range of overlap of $\Phi$ with arbitrary pure states, which may be interpreted to mean $\Phi$ represents a purer state. This is particularly clear in the two dimensional case given in \eq{2dpolarizationPhi}, where $\lambda_{\text{max}}=\lambda_1$ and $\lambda_{\text{min}}=\lambda_2$, increasing $\lambda_1$ and decreasing $\lambda_2$ in this case clearly leads to a state of higher polarization. 

Based on the above, we may postulate the following partial ordering on the three dimensional polarization states at hand. Suppose we have two arbitrary polarization states $\vec{\lambda}$ and $\vec{\mu}$, that satisfy the conditions
\begin{align}
\lambda_1 &\le \mu_1, \hspace{15pt} \lambda_3 \ge \mu_3. \label{eqmajcond}
\end{align}
We postulate that the conditions in \eq{eqmajcond} are equivalent to the statement that the state $\vec{\lambda}$ is less polarized than the state $\vec{\mu}$. This simple criterion has a deeper elegant explanation in the idea of majorization, which we proceed to define and whose significance for ordering polarization states we outline below.

Suppose we have two real vectors $\vec{x} \equiv (x_1, ..., x_N)$, and $\vec{y} \equiv (y_1, ..., y_N)$. We then define $\vec{x}^\downarrow$ and $\vec{y}^\downarrow$ as the same vectors with elements in descending order. For example, $x^\downarrow_1$ is the largest element in the vector $\vec{x}$. We say $\vec{x}$ majorizes $\vec{y}$, (i.e. $y$ is majorized by $\vec{x}$), written $\vec{x} \succ \vec{y}$, if we have
\begin{equation}
\sum_{i=1}^k x^\downarrow_i \ge \sum_{i=1}^k y^\downarrow_i, \hspace{10pt} k=1,..., N,
\label{majconditions}
\end{equation}
with equality holding when $k=N$. Majorization then provides a partial ordering on real vectors \cite{bhatia}.

For example, the vector $\vec{a} = (\frac{1}{2},\frac{1}{3},\frac{1}{6})$ majorizes the vector $\vec{b} = (\frac{2}{5},\frac{2}{5},\frac{1}{5})$. But the vector $\vec{c} = (\frac{3}{5},\frac{1}{5}, \frac{1}{5})$ niether majorizes nor is majorized by $\vec{a}$. Note that majorization works even when the elements of the two vectors do not sum to the same quantity, but in our examples they all sum to unity since this is required of the vectors of eigenvalues to which we apply majorization.

Majorization plays an important role in quantification of entanglement. It has been shown that for two bipartite quantum states $\ket{\psi}$ and $\ket{\phi}$, the former can be transformed to the latter using local operations and classical communication (LOCC) if and only if $\vec{\lambda}_\psi \prec \vec{\lambda}_\phi$, where the last two vectors represent the spectrum of the density matrix of the subsystems for each state \cite{nielsenmajorization}. Since LOCC transformations can never increase entanglement, this implies that majorization in the sense above is a partial order on entanglement. 

We recall the direct relationship between the amount of entanglement of a bipartite system with the purity of any one of its subsystems once the other subsystems have been traced out. More entanglement in the bipartite system directly means more mixedness (less purity) in the subsystem, the extremal cases being a maximally entangled bipartite state whose subsystem is maximally mixed (zero purity), and an unentangled (separable) bipartite state, whose subsystems are completely pure (purity 1). Since majorization is a partial ordering on degree of entanglement, the above implies it is also a partial ordering on quantum purity. Since quantum purity has been shown to be the same quantity mathematically as classical degree of polarization \cite{gamelpurity}, we can conclude the main premise of this paper: that majorization is a partial ordering on classical polarization.

In other words, we can conclude that for two arbitrary polarization states $\vec{\lambda}$ and $\vec{\mu}$,  if $\vec{\lambda} \prec \vec{\mu}$ then the polarization state $\vec{\lambda}$ is less polarized than than the state $\vec{\mu}$. Therefore, an admissible measure of polarization $P$ must satisfy $P( \vec{\lambda}) \le P(\vec{\mu})$ whenever $\vec{\lambda} \prec \vec{\mu}$. The previous statement is equivalent to stating that $P(\vec{\lambda})$ is a Schur-convex function \cite{schur}.

Applying the conditions in \eq{majconditions} we find that for $\vec{\lambda} \prec \vec{\mu}$ to hold, we must have
\begin{align}
\lambda_1 &\le \mu_1, \nonumber \\
\lambda_1 + \lambda_2 &\le \mu_1 + \mu_2, \label{eq2} \\
\lambda_1 + \lambda_2 + \lambda_3 &= \mu_1 + \mu_2 + \mu_3 \label{eqsum}.
\end{align}
The last \eq{eqsum} is always true for any polarization states X and Y, as the eigenvalues always sum to one. Taking \eq{eq2} and \eq{eqsum} we can find a condition on the third eigenvalues. Then, we have that $\vec{\lambda} \prec \vec{\mu}$ is equivalent to the conditions established in \eq{eqmajcond}.

Therefore, we have reduced the condition for majorization to conditions on the largest and smallest eigenvalues. 


\section{Polarization State Space}\label{s4:statespace}

\begin{figure}[htbp]
\centerline{\includegraphics[width=\columnwidth]{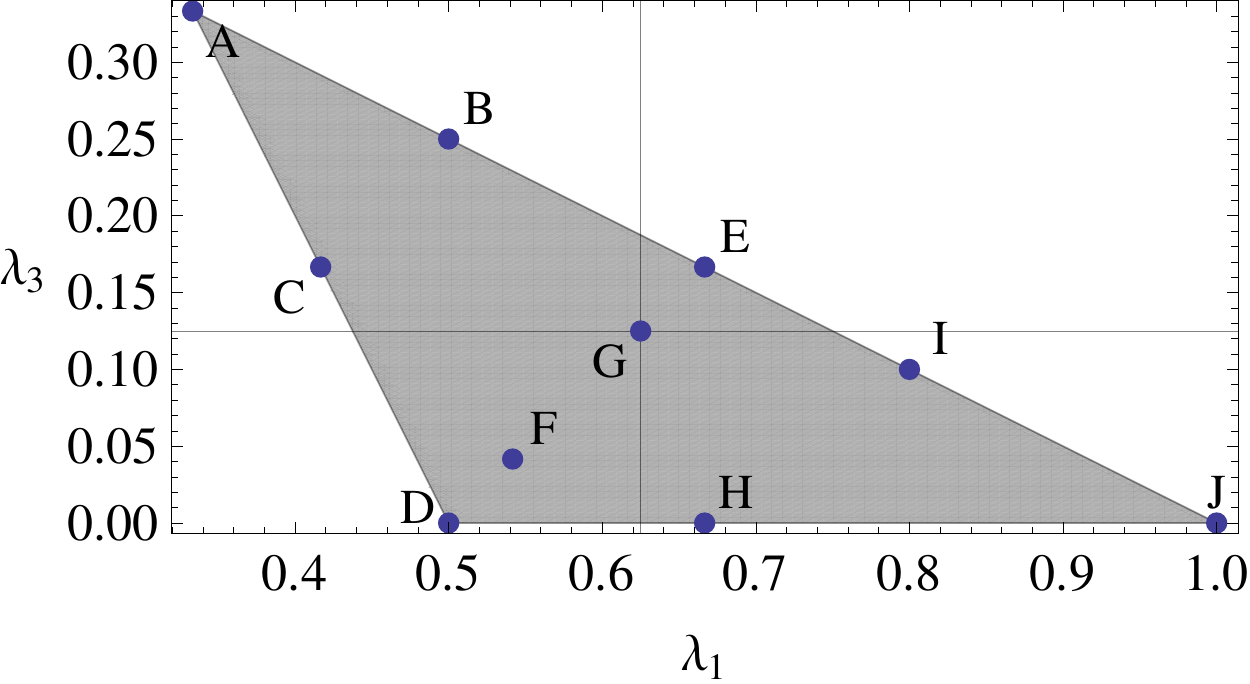}}
\caption{The shaded region represents the space of allowable maximum and minimum eigenvalues, $\lambda_1$ and $\lambda_3$, of the three dimensional polarization matrix. The middle eigenvalue is then given by $\lambda_2 = 1- \lambda_1 - \lambda_3$. The shaded region satisfies the inequalities in \eq{lambda3cond}, and non-negativity requirements. The labeled points A-J represent specific polarization states that are listed in table \ref{tableofstates} ahead. State A is the fully unpolarized state $(\frac{1}{3},\frac{1}{3},\frac{1}{3})$ and J the fully polarized state $(1, 0,0)$. To illustrate the majorization relations, we separate the area into quadrants centered around the state G given by $(\frac{5}{8},\frac{1}{4},\frac{1}{8})$.}
\label{fig1}
\end{figure}

Since we have introduced a partial ordering or polarization states that depends on the largest and smallest eigenvalues, $\lambda_1$ and $\lambda_3$ , it is of interest to find the allowed range of these eigenvalues for arbitrary polarization states. That is, we explore and graphically represent the state space. In the case of $\lambda_1$, we have the following condition
\begin{equation}
1 \ge \lambda_1 \ge \frac{1}{3}.
\label{lambda1cond}
\end{equation}
We also have 
\begin{equation}
2\lambda_1 + \lambda_3 \ge \lambda_1 + \lambda_2 + \lambda_3 = 1 \ge \lambda_1 + 2\lambda_3,
\label{rangeequalities}
\end{equation}
which for a given value of $\lambda_3$ leads to the following condition on $\lambda_3$ 
\begin{equation}
\frac{1}{2}(1-\lambda_1) \ge \lambda_3 \ge 1-2\lambda_1.
\label{lambda3cond}
\end{equation}
So the space of polarization states is the space of maximum and minimum (nonnegative) eigenvalues $\lambda_1$ and $\lambda_3$ that satisfy \eq{lambda1cond} and \eq{lambda3cond}. This is illustrated in fig. \ref{fig1}, where the shaded triangular region contains the allowable three dimensional polarization states. The figure includes ten points, each representing a polarization state outlined in table \ref{tableofstates}. We evaluate the degree of polarization of each state according to each of the seven measures of polarization in eqs. (\ref{pstandard} - \ref{pbarakat}). Note that for a given measure of polarization, the ordering it induces on the states is more important than the actual numerical degree of polarization assigned to a given state.

In the figure, we choose state G, and divide the state space to four quadrants around it, in order to illustrate the majorization relations in \eq{eqmajcond}. The states in the lower right quadrant, H,I and J, majorize (i.e. are more polarized than) state G. The states in the upper left quadrant, A, B, and C, are majorized by (i.e. are less polarized than) state G. 

The states in the other two quadrants, D, E, F, are neither majorized by, nor majorize state G. Therefore, they may be more or less polarized than G, and the majorization criterion tells us nothing about their relationship to G. In the following section, we develop another criterion that establishes a further partial ordering on the polarization state space.

As mentioned above, the generalization to three spatial dimensions of the Bloch sphere is an irregular shape within an Eight dimensional hypersphere \cite{kimura}. Yet each state in such a construction can be reduced, up to a diagonalizing unitary transformation, to a state in the shaded triangle in fig. \ref{fig1}. The state in our figure below should then uniquely determine all the basis independent properties of the polarization state.

\begin{table}[h]
 \begin{tabular}{c | c || c | c | c | c | c}  
 &  & $P_s$ & $P_v$ & $P_1$ & $P_2$ & $P_b$ \\ \hline
A & $(\frac{1}{3}, \frac{1}{3}, \frac{1}{3})$ & 0.333 &-1.58 & 0 & 0 & 0 \\ 
B & $(\frac{1}{2}, \frac{1}{4}, \frac{1}{4})$ & 0.375 & -1.5 & 0.25 & 0.25 & 0.395 \\ 
C & $(\frac{5}{12}, \frac{5}{12}, \frac{1}{6})$ & 0.375 & -1.48 & 0 & 0.25 & 0.468 \\ 
D & $(\frac{1}{2}, \frac{1}{2}, 0)$ & 0.5 & -1 & 0 & 0.5 & 1 \\ 
E & $(\frac{2}{3}, \frac{1}{6}, \frac{1}{6})$ & 0.5 & -1.25 & 0.5 & 0.5 & 0.707 \\ 
F & $(\frac{13}{24}, \frac{5}{12}, \frac{1}{24})$  & 0.469 & -1.20 & 0.125 & 0.5 & 0.864 \\ 
G & $(\frac{5}{8}, \frac{1}{4}, \frac{1}{8})$ & 0.469 & -1.30 & 0.375 & 0.5 & 0.688 \\ 
H & $(\frac{2}{3}, \frac{1}{3}, 0)$ & 0.556 & -0.918 & 0.333 & 0.667 & 1 \\ 
I & $(\frac{4}{5}, \frac{1}{10}, \frac{1}{10})$ & 0.66 & -0.922 & 0.7 & 0.7 & 0.885 \\ 
J & $(1,0,0)$  & 1 & 0 & 1 & 1 & 1
\end{tabular}

\caption{\label{tableofstates} The five independent measures of polarization defined in  eqs. (\ref{pstandard}) and (\ref{pvonn} - \ref{pbarakat}) evaluated for ten polarization states of interest.
} 
\end{table}

If we compare the states in table \ref{tableofstates}, we find we can give a partial order on them through majorization as per
\begin{align}
\vec{\lambda}^A \prec \vec{\lambda}^B, \vec{\lambda}^C \prec \vec{\lambda}^D, & \vec{\lambda}^E, \vec{\lambda}^F, \vec{\lambda}^G \prec \vec{\lambda}^H \prec \vec{\lambda}^J,\nonumber\\
\vec{\lambda}^E, \vec{\lambda}^G &\prec \vec{\lambda}^I \prec \vec{\lambda}^J.
\label{majorder}
\end{align}
Therefore, we require that our chosen measure of polarization respect the partial ordering of states induced by majorization, i.e. that it be a Schur-convex function.

\section{Testing the Majorization Criterion}\label{s5:majtest}

Examining table \ref{tableofstates}, we find that the measures $P_1$ and $P_b$ violate the ordering due to majorization in \eq{majorder}. For example, $P_1$ ranks the state H as less polarized than G while the majorization criterion dictates that the opposite should be the case. Similarly, $P_b$ gives the same polarization to states D and J, while J majorizes D and should therefore have a higher polarization.

The other measures respect the majorization criterion for the states above. However, we still need to prove that they will always respect this criterion for any state. 

Suppose once more that we have two distinct polarization states $\vec{\lambda}$ and $\vec{\mu}$ that satisfy $\vec{\lambda} \prec \vec{\mu}$. They can always be written as
\begin{align}
\vec{\lambda} &= (\lambda_1, \lambda_2, \lambda_3), \nonumber \\
\vec{\mu} &= (\lambda_1 {+} a, \lambda_2 {-} a {+} b, \lambda_3 {-} b),
\label{comparewxy}
\end{align}
for some $a, b \ge 0$. Given \eq{comparewxy}, we then verify that $P_s$, $P_v$ and $P_2$ will always assign a higher polarization to $\vec{\mu}$ than $\vec{\lambda}$. In case of $P_s$, we have
\begin{align}
&P_s(\vec{\mu}) - P_s(\vec{\lambda}) \nonumber\\
&= [(\lambda_1 {+} a)^2 + (\lambda_2 {-} a {+} b)^2 +  (\lambda_3 {-} b)^2] - ( \lambda_1^2 + \lambda_2^2 + \lambda_3^2) \nonumber\\
&= 2[a\lambda_1 {+} (b{-}a)\lambda_2 {-}b\lambda_3] + a^2 + b^2 + (b{-}a)^2 \nonumber\\
&= 2[a(2\lambda_1 {+} \lambda_3 {-} 1) + b(1{-}\lambda_1 {-} 2\lambda_3)] + a^2 + b^2 + (b{-}a)^2  \nonumber\\
&\ge0,
\label{pscriteria}
\end{align}
where in the third line we made use of $\lambda_2 = 1-\lambda_1-\lambda_3$. Also note that the two terms inside the square brackets on the third line are non-negative by \eq{rangeequalities}, which leads to the non-negativity of the whole expression. We have show in \eq{pscriteria} that $P_s$ (and therefore $P_{bl}$ and $P_{sc}$) always respects the majorization criterion.

In the case of $P_2$, we have
\begin{align}
P_2(\vec{\mu}) {-} P_2(\vec{\lambda}) &= [(\lambda_1{+} a) - (\lambda_3 {-} b)] - (\lambda_1{-} \lambda_3) \nonumber\\
&= a+ b \ge 0.
\label{p2criteria}
\end{align}
Therefore, by \eq{p2criteria}, the measure $P_2$ respects the majorization criterion. 

We use a slightly different approach for $P_v$. First, we define a function equal to the measure, but takes only the largest and smallest eigenvalues as inputs. Define
\begin{align}
f_v(\lambda_1, \lambda_3) \equiv &\lambda_1 \log_2\lambda_1 + (1{-}\lambda_1{-}\lambda_3) \log_2(1{-}\lambda_1{-}\lambda_3)  \nonumber\\
&+ \lambda_3 \log_2\lambda_3.
\end{align}
Then we find the partial derivatives
\begin{align}
\frac{\partial f_v}{\partial \lambda_1} = \log_2\lambda_1 - \log_2(1{-}\lambda_1{-}\lambda_3),\nonumber\\
\frac{\partial f_v}{\partial \lambda_3} = \log_2\lambda_3 - \log_2(1{-}\lambda_1{-}\lambda_3).
\label{fpartials}
\end{align}
We are interesting in the points where these partial derivatives are zero. We find that $\frac{\partial f_v}{\partial \lambda_1}=0$ only when $2\lambda_1 {+} \lambda_3 {-} 1 = 0$, which turns out to coincide with the line AD bordering the allowable region. Similarly, $\frac{\partial f_v}{\partial \lambda_1}=0$ only when $1{-}\lambda_1 {-} 2\lambda_3=0$, which coincides with the line AJ. 

Therefore, we find that the partial derivatives of this measure only take a value zero on the edges of the allowable region given by \eq{rangeequalities}. For all points within the allowable region, we have from \eq{fpartials} that $\frac{\partial f_v}{\partial \lambda_1}>0$ and $\frac{\partial f_v}{\partial \lambda_3}<0$. This implies that increasing $\lambda_1$ will always increase the von Neumann degree of polarization $P_v$, as will decreasing $\lambda_3$, that satisfying the majorization criterion. 

\section{Graphical Comparison}\label{s6:graphcomp}

\begin{figure}
\centerline{\includegraphics[width=\columnwidth]{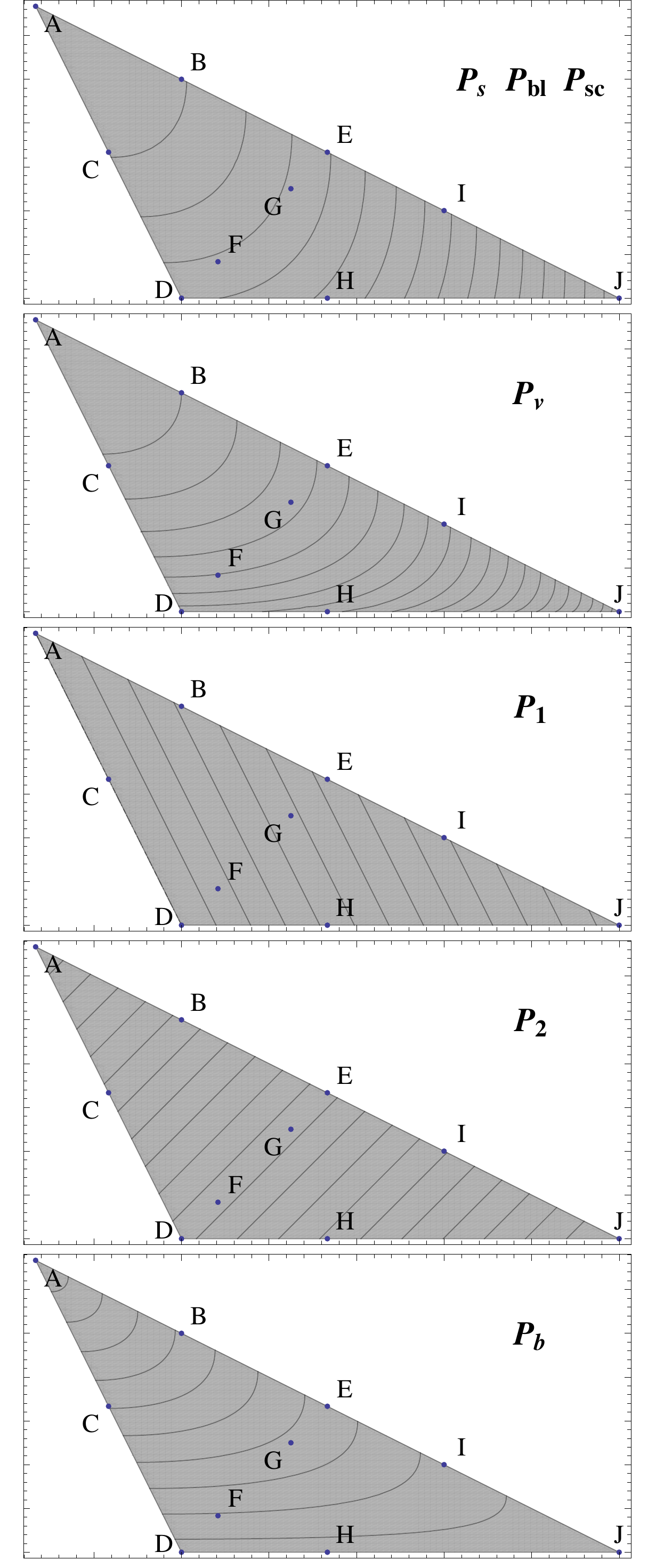}}
\caption{Graphs of the state space with contours of constant degree of polarization for each of the measures under consideration. Each graph is essentially fig. \ref{fig1} with the contours of constant polarization superimposed. The measure used is in the top right of each diagram. The points A-J defined in table \ref{tableofstates} and plotted in previous figures are shown in the graphs. The measures $P_s$, $P_{bl}$ and $P_{sc}$ have the same contours since they are functions of one another.
}
\label{figcontours}
\end{figure}

To get more insight, we compare all the measures graphically in fig. \ref{figcontours}. In the figure, we plot the contours of equal polarization in the region of allowable physical states, for each measure of polarization. 

We observe that the contours of $P_1$ are straight lines with a negative slope, clearly violating the majorization criterion between any two points on the same contour line. The measure $P_b$ violates the majorization criterion as well since it has a contour that is a straight line connecting DJ. 

We also note that the contours of $P_s$, $P_v$ and $P_2$ in the three other figures never have a negative slope, and may have a zero or infinite slope only at isolated points at the edge of the allowable region. Therefore as shown in the equations above, they satisfy the majorization criterion we have established. So we are left to choose a measure from this list.

We note that the contours of $P_s$ and $P_2$ have a certain interesting symmetry lacking in $P_v$, they are invariant in reflections about the line AH. This reflection maps together the pairs of points B and C, D and E, F and G, and leaves the contours for these two measures unchanged. The measure $P_v$ also has contours that are very nearly horizontal for relatively large ranges, meaning it is less sensitive to changes in the state for these values. So we choose to discard $P_v$.

The measure $P_2$ seems adequate and is very simple to calculate, as seen from its straight contours, and will suffice for most purposes. However, if we go back to the way it was defined, we notice that it oversimplifies the situation. Both $P_1$ and $P_2$ were defined by first writing the polarization matrix as a positive linear combination of rank 1, 2, and 3 matrices. The rank 1 matrix is clearly fully polarized, while the rank 3 is completely unpolarized. A reasonable intuitive assessment suggests that the rank 2 matrix should have an intermediate partial polarization, between the other two. 

The weakness of the $P_1$ measure is that it ignores the partial polarization in the rank 2 component, and treats it as if it were completely unpolarized. We see that $P_2$ has the opposite weakness, in that treats the rank 2 component the same way as the rank 1 component, i.e. it treats it as if it were fully polarized. Both approaches are inadequate. A suitable measure is one that gives a rank 2 matrix an intermediate polarization between rank 1 and rank 3. 

So we are left with $P_s$ and its closely related measures $P_{bl}$ and $P_{sc}$. Since they all induce the same ordering on polarization states, the difference between them is immaterial for most uses. However, $P_{bl}$ defined in \eq{pbloch} seems the most physically motivated and mathematically elegant. It also has some interesting properties in the scaling of polarization if a depolarizing channel is applied to the field \cite{gamelpurity}.

\section{Generalization to Higher Dimensions}\label{s7:Ndimensions}
The results we have obtained hitherto in the paper are specifically tailored to three spatial dimensions. However most of them can easily be generalized to higher dimensions in a straightforward manner. We consider the optical polarization in $N$ spatial dimensions, or the mathematically equivalent quantum purity of an $N$-level quantum system. 

The polarization matrix $\Phi$ is assumed to be $N\times N$ in dimension, and have eigenvalues $\lambda_1 \ge ... \ge \lambda_N \ge 0$, that sum to unity. The measures of polarization $P_s$, $P_{bl}$, $P_{sc}$, $P_v$, and $P_b$ are then given by
\begin{align}
P_s &= \sum_{i=1}^N \lambda_i^2 \equiv \Tr[\Phi^2], \\
P_{bl} &= \sqrt{\frac{NP_s - 1}{N-1}}, \\
P_{sc} &= \frac{1}{P_s}, \\
%
P_v &= \sum_{i=1}^N  \lambda_i \log_2{ \lambda_i},\\
P_b &= \sqrt{1 - N^N\lambda_1\lambda_2...\lambda_N}.
\end{align}

We will also have a hierarchy of measures that generalizes $P_1$ and $P_2$. The diagonalization of $\Phi$ can be written as a unique positive sum of $N$ matrices, of rank $1, 2 ... N$. The sum of the first $k$ of these coefficients can be taken as a measure of polarization. We will then have
\begin{equation}
P_k = \lambda_1 - \lambda_{k+1}, \hspace{20pt} k=1, ... , N-1.
\end{equation}

Moreover, there will be a hierarchy of $N-1$ measures due to Barakat \cite{barakat83}, the first of which will always be the Bloch measure $P_{bl}$, and the last of which, by definition will be $P_b$, Barakat's last measure.

The majorization criterion can still be used in $N$ dimensions by applying \eq{majconditions} to the vector of eigenvalues. However, for $N>3$, the result will not be as neat as the simple inequalities in \eq{eqmajcond}. 

One can choose $N-1$ independent variables (that are linear combinations of the eigenvalues) to represent the state. The region of allowable states will consist of a polytope in $N-1$ dimensions that generalizes the shaded triangular region shown in our figures.

Even in $N$ dimensions, the measure $P_{bl}$ remains the most useful in the large majority of cases for the same reasons discussed earlier in the paper.

\section{Conclusion}
We have recapped the most prevalent measures of three dimensional polarization in the literature. We then represented the space of allowable polarization states in simple geometric terms, as a function of the highest and lowest eigenvalues. Furthermore, we introduced the criterion of majorization provides a partial ordering and an equivalence relation between the degree of polarization of states in the state space. 

We showed which of the measures are consistent with this criterion, and represented all the measures graphically, identifying interesting properties of each. Based on masjorization and graphical additional considerations, we chose the Bloch measure, $P_{bl}$ of polarization as the most versatile of the measures of polarization. We also demonstrated that much of the analysis can be generalized to $N$ dimensions.


\section*{Acknowledgements}
This work was funded by the Natural Sciences and Engineering Research Council of Canada (NSERC).


\begin{thebibliography}{99}

\bibitem{stokes}
G.~G. Stokes.
\newblock On the composition and resolution of streams of polarized light from
  different sources.
\newblock {\em Trans. Cambridge Philos. Soc.}, 9:399--416, 1852.

\bibitem{poincare}
H.~Poincar\'{e}.
\newblock Leçons sur la théorie mathématique de la lumière ({Lectures} on
  the mathematical theory of light).
\newblock {\em G. Carr\'{e}. Paris}, IV--408, 1889.

\bibitem{wolf07}
E.~Wolf.
\newblock {\em Introduction to the theory of coherence and polarization of
  light}.
\newblock Cambridge, 2007.

\bibitem{qasimiwolf}
Asma Al-Qasimi, Olga Korotkova, Daniel James, and Emil Wolf.
\newblock Definitions of the degree of polarization of a light beam.
\newblock {\em Opt. Lett.}, 32(9):1015--1016, May 2007.

\bibitem{nielsenchuang}
M.~Nielsen and I.~Chuang.
\newblock {\em Quantum computation and quantum information}.
\newblock Cambridge, 2000.

\bibitem{bergman}
T.~Carozzi, R.~Karlsson, and J.~Bergman.
\newblock Parameters characterizing electromagnetic wave polarization.
\newblock {\em Phys. Rev. E}, 61:2024--2028, 2000.

\bibitem{friberg02}
T.~Set\"{a}l\"{a}, A.~Shevchenko, M.~Kaivola, and A.~T. Friberg.
\newblock Degree of polarization for optical near fields.
\newblock {\em Phys. Rev. E}, 66:016615, 2002.

\bibitem{qianeberly}
X.~F. Qian and J.~H. Eberly.
\newblock Entanglement and classical polarization states.
\newblock {\em Optics Letters}, 36(20):4110--4112, 2011.

\bibitem{gil07}
J.~J. Gil.
\newblock Polarimetric characterization of light and media.
\newblock {\em The European Physical Journal - Applied Physics}, 40:1--47, 10
  2007.

\bibitem{wolf04}
J.~Ellis, A.~Dogariu, S.~Ponomarenko, and E.~Wolf.
\newblock Degree of polarization of statistically stationary electromagnetic
  fields.
\newblock {\em Optics Communications}, 248(4--6):333--337, 2005.

\bibitem{barakat83}
R.~Barakat.
\newblock N--fold polarization measures and associated thermodynamic entropy of
  {N} partially coherent pencils of radiation.
\newblock {\em J. Mod. Opt.}, 30(8):1171--1182, 1983.

\bibitem{gamelpurity}
O.~Gamel and D.~F.~V. James.
\newblock Measures of quantum state purity and classical degree of
  polarization.
\newblock {\em Phys. Rev. A}, 86(3):033830, 2012.
\newblock arXiv:1303.6696.

\bibitem{nielsenmajorization}
M.~A. Nielsen.
\newblock Conditions for a class of entanglement transformations.
\newblock {\em Phys. Rev. Lett.}, 83:436--439, Jul 1999.

\bibitem{luis05}
A.~Luis.
\newblock Polarization distribution and degree of polarization for
  three--dimensional quantum light fields.
\newblock {\em Phys. Rev. A}, 71:063815, 2005.

\bibitem{friberg09}
T.~Set\"{a}l\"{a}, K.~Lindfors, and A.~T. Friberg.
\newblock Degree of polarization in {3D} optical fields generated from a
  partially polarized plane wave.
\newblock {\em Optics Letters}, 34(21):3394--3396, 2009.

\bibitem{friberg10}
T.~Voipio, T.~Set\"{a}l\"{a}, A.~Shevchenko, and A.~T. Friberg.
\newblock Polarization dynamics and polarization time of random
  three--dimensional electromagnetic fields.
\newblock {\em Phys. Rev. A}, 82:063807, 2010.

\bibitem{sheppard}
C.~J.~R. Sheppard.
\newblock Geometric representation for partial polarization in three
  dimensions.
\newblock {\em Optics Letters}, 37(14):2772--2774, 2012.

\bibitem{gil10}
J.~J. Gil and I.~San~Jos{\'e}.
\newblock 3d polarimetric purity.
\newblock {\em Optics Communications}, 283:4430--4434, 2010.

\bibitem{gellmann}
M.~Gell-Mann and Y.~Ne'eman.
\newblock {\em The Eightfold Way}.
\newblock Benjamin, New York, 1964.

\bibitem{bloch}
F.~Bloch.
\newblock Nuclear induction.
\newblock {\em Phys. Rev.}, 70(7--8):460--474, 1946.

\bibitem{kimura}
G.~Kimura.
\newblock {The Bloch vector for N--level systems}.
\newblock {\em Phys. Lett. A}, 314(5--6):339--349, 2003.

\bibitem{schmidt}
E.~Schmidt.
\newblock Zur theorie der linearen und nichtlinearen integralgleichungen. {III.
  Teil}.
\newblock {\em Mathematische Annalen}, 65(3):370--399, 1908.

\bibitem{shannon}
C.~E. Shannon.
\newblock A mathematical theory of communication.
\newblock {\em Bell System Technical Journal}, 27:379--423, 623--656, 1948.

\bibitem{bhatia}
R.~Bhatia.
\newblock {\em Matrix Analysis}.
\newblock Graduate Texts in Mathematics. Springer New York, 1997.

\bibitem{schur}
AlbertW. Marshall, Ingram Olkin, and BarryC. Arnold.
\newblock Schur-convex functions.
\newblock In {\em Inequalities: Theory of Majorization and Its Applications},
  Springer Series in Statistics, pages 79--154. Springer New York, 2011.

\end{thebibliography}

\end{document}